\documentclass[reprint,amsmath,amssymb,aps,prl,showpacs]{revtex4-1}
\usepackage[utf8]{inputenc}
\usepackage{graphicx}
\usepackage[usenames,dvipsnames]{color}
\usepackage{dcolumn}
\usepackage{bm}
\usepackage[breaklinks]{hyperref}
\usepackage{bbold}
\usepackage[normalem]{ulem}
\usepackage{cancel}
\usepackage{gensymb}
\usepackage{times}
\usepackage{booktabs}

\renewcommand*{\HyperDestNameFilter}[1]{\jobname-#1}

\renewcommand{\vec}[1]{\mathbf{#1}}

\newcommand{\kk}{{\vec k}_0}

\newcommand{\citationtitle}[1]{{\color{Gray}\small #1}}

\graphicspath{{Figures/}}

\begin{document}

\title{Spin-Peierls Instability of Three-Dimensional Spin Liquids with Majorana Fermi Surfaces}

\author{Maria Hermanns}
\author{Simon Trebst}
\author{Achim Rosch}
\affiliation{Institute for Theoretical Physics, University of Cologne, 50937 Cologne, Germany}

\date{\today}

\begin{abstract}
Three-dimensional (3D) variants of the Kitaev model can harbor gapless spin liquids with a Majorana Fermi surface on certain tricoordinated lattice structures such as the recently introduced hyperoctagon lattice.
Here we investigate Fermi surface instabilities arising from additional spin exchange terms (such as a Heisenberg coupling) which introduce interactions between the emergent Majorana fermion degrees of freedom.
We show that independent of the sign and structure of the interactions, the Majorana surface is always unstable.
Generically  the system spontaneously doubles its unit cell  at exponentially small temperatures and forms a spin liquid with  line nodes. 
Depending on the microscopics  
further symmetries of the system can be broken at this transition.
These spin-Peierls instabilities of a 3D spin liquid are closely related to  BCS instabilities of fermions. 
\end{abstract}

\maketitle

The interplay of frustration and correlations in quantum magnets engenders a rich variety of quantum states with fractional excitations that are collectively referred to as quantum spin liquids (QSL) \cite{SpinLiquids}. Archetypal instances of such states include gapped topological QSLs with anyonic quasiparticle excitations 
as well as gapless QSLs 
where emergent spinon excitations form nodal structures such as Dirac points, Fermi lines or Fermi surfaces reminiscent of metallic states. A family of such gapless QSLs is realized in two- and three-dimensional variants of the Kitaev model \cite{kitaev2006}, in which a high level of exchange-frustration is induced by competing bond-directional interactions of the form
\begin{equation}
  H_{\rm Kitaev} = J \sum_{\gamma \rm-links} \sigma_i^\gamma \sigma_j^\gamma \,.
  \label{eq:Kitaev}
\end{equation}
Here spin-1/2 moments $\sigma$ on sites $i$ and $j$ are coupled via an Ising-like exchange whose easy-axis aligns with the $\gamma = x,y,z$ orientation of the bonds of the underlying tricoordinated lattices. Kitaev's seminal solution \cite{kitaev2006} of this model for the honeycomb lattice allows to analytically track the fractionalization of the original spin-1/2 moments into emergent massless Majorana fermions (spinons) and massive $\mathbb Z_2$ gauge excitations (visons). The collective QSL ground state is a (semi)metallic state formed by the itinerant Majorana fermions. The qualitative nature of this Majorana metal turns out to depend on  the underlying lattice: for the two-dimensional honeycomb lattice the itinerant Majorana fermions form two Dirac cones \cite{kitaev2006} (akin to the well-known electronic band structure of graphene), while for the three-dimensional hyperhoneycomb  and hyperoctagon lattices the gapless Majorana modes form  a Fermi line \cite{Mandal09} and an entire Fermi surface \cite{hermanns14}, respectively. In the presence of additional time-reversal symmetry breaking terms the Majorana fermions (on the hyperhoneycomb lattice) can even form a topological  semimetal with Weyl nodes \cite{wsl2014}.
Interest in such three-dimensional Kitaev models has been sparked by the recent experimental observation of spin-orbit entangled $j=1/2$ Mott insulators with strong bond-directional interactions of the form \eqref{eq:Kitaev} in the iridates $(\beta,\gamma)$-Li$_2$IrO$_3$ \cite{beta,gamma,Biffin,Coldea} , where the iridium sites form three-dimensional, tricoordinated lattice structures. The synthesis of such three-dimensional Kitaev structures expands an intense ongoing search for solid-state realizations of the original two-dimensional Kitaev model, which -- following the early theoretical guidance of Khaliullin and coworkers \cite{Khaliullin} -- has put materials such the layered iridates Na$_2$IrO$_3$, $\alpha$-Li$_2$IrO$_3$ \cite{honeycomb-iridates} and more recently $\alpha$-RuCl$_3$ \cite{RuCl3} into focus.

In this paper, we will concentrate on  three-dimensional Kitaev spin liquids characterized by Majorana Fermi surfaces and show that they generically dimerize, i.e. double their unit cell at low temperature. This instability can be viewed as a generalization of the spin-Peierls transition in one-dimensional systems \cite{SpinPeierls1,SpinPeierls2}, or more generally, of the tendency of frustrated low-dimensional spin systems to form valence-bond solids \cite{Sachdev}. The spin-Peierls instability of (quasi-) one-dimensional spin systems describes that an arbitrarily small coupling of a spin chain to classical lattice degrees of freedom leads at  low temperature to a dimerization and a gap in the spin system \cite{SpinPeierls1,SpinPeierls2}: the energy gain by opening the gap is larger than the energy needed to distort the lattice. When the phonon mode is, however, treated quantum mechanically, a dimerization occurs only when a critical coupling strength is reached \cite{Sandvik}. 
As we will show, for the 3D Kitaev spin liquid a variant of the spin-Peierls transition occurs at low $T$ even in the {\em absence} of lattice degrees of freedom and for arbitrarily weak perturbations. Notably, the result of this instability is not a short-range valence-bond ordered state, but still a QSL -- yet one, in which the original Majorana Fermi surface has collapsed into a line of gapless modes.

\begin{figure}[t]
\begin{center}
\includegraphics[width=0.9 \linewidth]{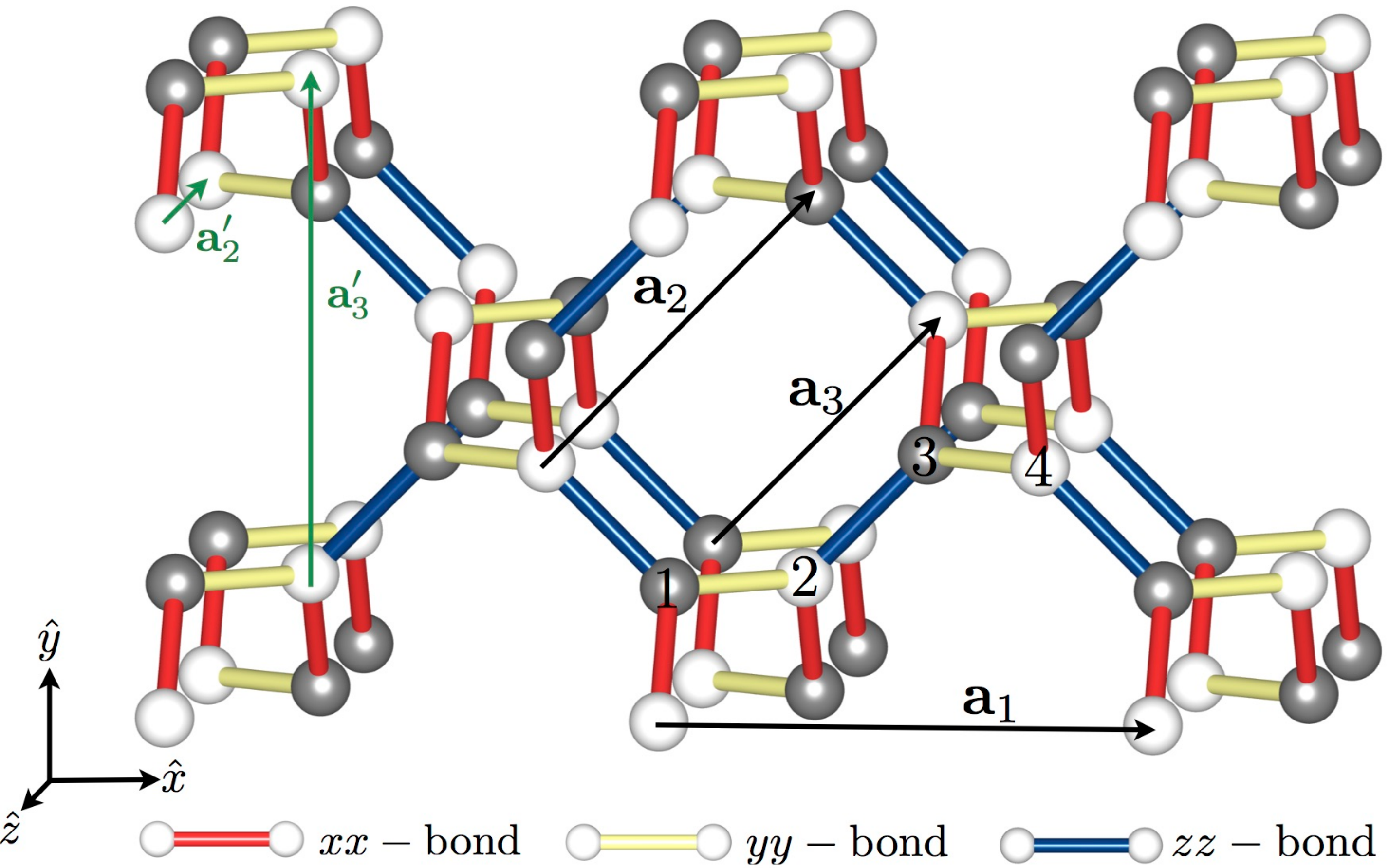}
\end{center}
\caption{(color online) The hyperoctagon lattice is a tricoordinated bcc lattice without inversion symmetry (space group I4$_1$32) and a four-site unit cell (with atom positions  $1,\dots,4$ and unit cell vectors $\vec a_1$, $\vec a_2$ and $\vec a_3$ indicated in the figure). For the definition of the Kitaev model on this lattice we define nearest neighbor spin interactions of the $\sigma^x\sigma^x$, $\sigma^y\sigma^y$, and $\sigma^z\sigma^z$ type assigned to the green, red, and blue bonds, respectively. The vectors $\vec a_1$, $\vec a_2'=\vec a_2-\vec a_3$, and $\vec a_3'=-\vec a_1 +\vec a_2 +\vec a_3$ are the lattice translation vectors of each of the two sublattices (denoted by white and grey lattice sites). }
\label{fig:hyperoctagon}
\end{figure}

\noindent {\em Model system.--} 
For concreteness, we will focus our analysis on a specific 3D Kitaev model and return to a more general discussion later. 
We look at the Kitaev model defined on the so-called hyperoctagon lattice depicted in Fig.~\ref{fig:hyperoctagon}. For this system a gapless QSL ground state with a Majorana Fermi surface was established in Ref.~\onlinecite{hermanns14}, to which we also refer for technical details of the analytical solution of this model.
In brief, the spin degrees of freedom fractionalize to itinerant Majorana fermions interacting with static $\mathbb{Z}_2$ gauge fields \cite{kitaev2006}. 
In the (flux-free) ground state sector of the $\mathbb{Z}_2$ gauge field, the original spin model \eqref{eq:Kitaev} reduces to a free hopping model of Majorana fermions
 \begin{align}
H_0&=-i J \sum_{\mathbf{R}} 
c_{1}(\mathbf{R})c_{3}(\mathbf{R}-\mathbf{a}_{2}) 
+ c_{2}(\mathbf{R})c_{4}(\mathbf{R}-\mathbf{a}_{3}) \nonumber\\
&+ c_{1}(\mathbf{R})c_{2}(\mathbf{R})
- c_{3}(\mathbf{R})c_{4}(\mathbf{R})
- c_{2}(\mathbf{R})c_{3}(\mathbf{R}) \nonumber\\
&+ 
c_{1}(\mathbf{R})c_ {4}(\mathbf{R}-\mathbf{a}_{1}) \,.
\label{eq:hyperoctagon}
\end{align} 
Here $\vec R$ is the position of the unit cells within the bcc Bravais lattice (see Fig.~\ref{fig:hyperoctagon}), each containing four sites labeled by $i=1,2,3,4$; $\vec a_j$ are the lattice vectors defined in Fig.~\ref{fig:hyperoctagon} and the Majorana operators obey the usual  algebra $\{c_{i}(\vec R),c_j(\vec R')\}=2 \delta_{ij} \delta_{\vec R, \vec R'}$.

\begin{figure}[t]
\begin{center}
\includegraphics[width=\linewidth]{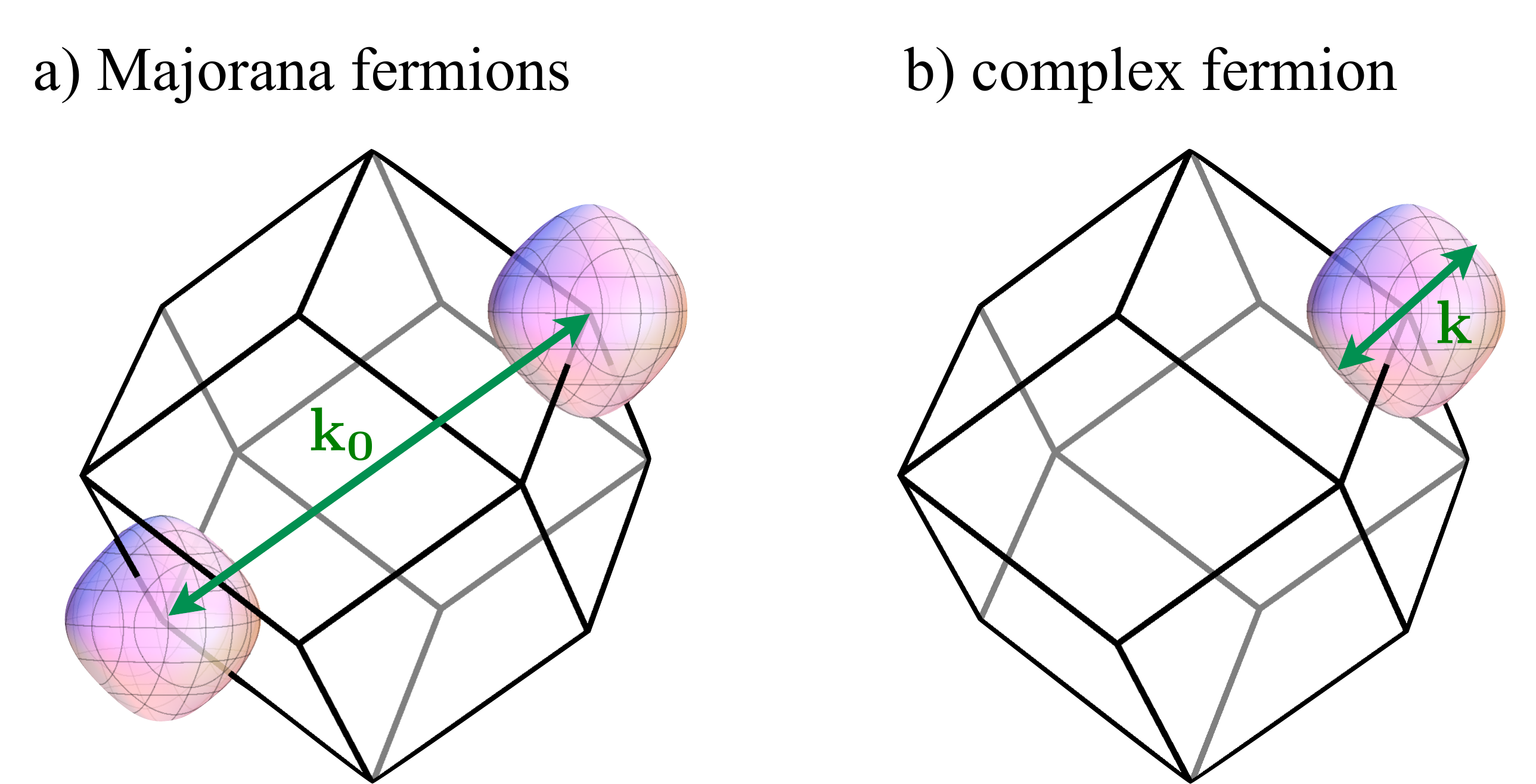}
\end{center}
\caption{(color online) a) Sketch of the two Majorana Fermi surfaces in the Brillouin zone. In the Majorana language, the Peierls instability is due to perfect nesting between the two surfaces.  b) The two Majorana surfaces combine to a single Fermi surface of a (complex) fermion. In the fermionic language, the Peierls instability is the usual BCS instability.}
\label{fig:surface}
\end{figure}

Diagonalizing the Hamiltonian \eqref{eq:hyperoctagon}, one finds  that two of the four bands have a gap of order $J$; the other two bands become gapless on two two-dimensional surfaces in momentum space defined by the equation $\cos k_x+\cos k_y + \cos k_z =- \frac{3}{2}$ \cite{hermanns14}. 
The two closed Majorana surfaces are centered around the
momenta $\pm \kk/2$ with $\kk=2 \pi (1,1,1)$ as illustrated in Fig.~\ref{fig:surface}~a). Note that only $2 \kk$ is a unit vector of the reciprocal lattice, but not $\kk$ itself.
Most importantly, there is a perfect nesting condition between the two Majorana surfaces, 
\begin{equation}
  \epsilon_{\vec k}=\epsilon_{\vec{k}+\kk} \,,
\end{equation}
 which is {\em not} specific to the microscopic Hamiltonian (\ref{eq:Kitaev}) but arises from the peculiar form of {\em time-reversal symmetry} $\mathcal T$ of the underlying spin model realized by \cite{kitaev2006,hermanns14} 
\begin{equation}
  c_j(\vec R) \stackrel{\mathcal T}{\to }  (-1)^j e^{i\vec k_0 \cdot \vec R} c_j(\vec R) \,.
  \label{eq:T}
\end{equation}

For the following discussion it is convenient to combine the two low-energy Majorana bands into a single band of a (complex) fermion, which allows to use the more familiar language of 
superconductors to describe the system. 
To do so, we denote the low-energy Majorana excitations with momentum $\vec k$  by $\gamma^{1/2}_{\vec k}$ with $\epsilon^1_{\vec k}\le \epsilon^2_{\vec k}$. Note that the particle-hole symmetry of Majorana fermions readily implies $\gamma_1(\vec k)=\gamma_2^\dagger(-\vec k)$, which requires to either restrict the discussion to half the Brillouin zone or, alternatively, to the upper energy band.
Doing the latter, we introduce the Fermion operator $f_{\vec k}$ by $f_{\vec k}=\frac{1}{\sqrt{2}} \gamma^2_{\vec k}$. 
This reformulation combines the two Majorana surfaces in Fig.~\ref{fig:surface} a) into a single Fermi surface of a complex fermion centered around $\kk/2$, see Fig.~\ref{fig:surface}~b).
In terms of the complex fermion, time-reversal symmetry  \eqref{eq:T} becomes
\begin{equation}
f^\dagger_{\vec k_0/2 + \vec k} \stackrel{\mathcal T}{\longrightarrow } f^\dagger_{\vec k_0/2 - \vec k}  \label{Tsym} \,,
\end{equation}
constraining the fermionic energy spectrum to be symmetric relative to $\vec k_0/2$: 
\begin{align}
E_{\vec k_0/2 +\vec k}= E_{\vec k_0/2 -\vec k} \,.
\label{eq:TRenergy}
\end{align} 
This energy relation naturally leads to consider pairing terms of the form $f^\dagger_{\vec k_0/2+\vec k}f^\dagger_{\vec k_0/2-\vec k}$. However, such pairs carry a {\em finite} momentum $\vec k_0$ \cite{FootnotePairing} and can therefore only arise in a phase in which translational symmetry is spontaneously broken. In the following, we will show that such a spontaneous symmetry breaking will occur in the presence of additional spin exchange terms.

\noindent {\em Interactions.--}
Any deviation from the ideal Kitaev model, e.g. by introducing spin-spin interactions of the Heisenberg form, 
Dzyaloshinskii Moriya interactions, three-spin interactions, etc., will induce interactions of the Majorana fermions. As long as the size of these perturbations is sufficiently small (compared to the Kitaev interaction $J$, which in particular  sets the energy scale for the flux  gap), the interaction can be written in terms of the Majorana fermions only (without any contributions from the fluxes) and will remain short-ranged. For the hyperoctagon lattice, two types of Majorana interactions turn out to be the most local ones, which are therefore also expected to be the most important ones in an expansion around the Kitaev model \cite{FootnoteThirdTerm}. We parametrize them by their overall strength $U$ and an angle $\alpha$ as
\begin{align}
H_{\rm int} &= - U  \Big(\cos\alpha \sum_{\vec R} c_1(\vec{R}) c_2(\vec{R}) c_3(\vec{R}) c_1(\vec{R}+\vec a_2) \nonumber\\
& + \sin \alpha  \sum_{\vec R} c_1(\vec{R}) c_2(\vec{R}) c_3(\vec{R}) c_4(\vec{R})+{\rm sym.} \Big) \label{int} \,.
\end{align}
Note that each term in the above expression is only a single representative of 24 distinct terms
obtained by the 48 symmetry transformations of the I4$_1$32 symmetry group of the hyperoctagon lattice
\cite{FootnoteSymmetryTransformation}.  
When adding, e.g., a nearest-neighbor Heisenberg coupling $J_H$ to the Kitaev model, one expects $U \sim J_H^{4}/J^{3}$  \cite{kitaev2006}.
In general, the sizes of $U$ and $\alpha$ will depend on all types of microscopic details and are difficult to predict quantitatively. We therefore analyze the effect of the interactions \eqref{int} for varying values of $\alpha$ (and fixed, small $U$).

To analyze $H_0+H_{\rm int} $ for small $U$, we first project the interaction \eqref{int} onto the 
low-energy degrees of freedom described by the complex $f$ fermions introduced above and obtain
\begin{align}
H_{\rm eff}=&\sum_{\vec k} \epsilon_{\vec  k} f^\dagger_{\bf k} f^{\phantom\dagger}_{\bf k} + \! \sum_{\vec k_1, \vec  k_2, \vec  k_3}\!\!
V^1_{\vec k_1, \vec k_2,\vec k_3} \,\, f^\dagger_{\vec k_1} f^\dagger_{\vec k_2} f^{\phantom\dagger}_{\vec k_3} f^{\phantom\dagger}_{\vec k_1 +\vec k_2 -\vec k_3}\nonumber \\
&+ \! \sum_{\vec k_1,\vec  k_2,\vec  k_3}\!\!
V^2_{\vec k_1, \vec  k_2,\vec  k_3}\,\,  f^\dagger_{\vec k_1} f^\dagger_{\vec k_2} f^\dagger_{\vec k_3} f^\dagger_{-\vec k_1 -\vec k_2 -\vec k_3}+h.c. \label{Hlow}
\end{align}
The momentum dependence of $V^{1,2}$ can easily be obtained numerically using Eq.~(\ref{int}), the eigenmodes of $H_0$,  and the definition $f_\vec{k}^\dagger$ given above.
Note that symmetry-allowed terms of the form $f^\dagger_{\vec k}f^\dagger_{-\vec k}$ or  $f^\dagger_{\vec k_1}f^\dagger_{\vec k_2}f^\dagger_{\vec k_3}f^{\phantom\dagger}_{\vec k_1+\vec k_2+\vec k_3}$ do not contribute to the low-energy sector as at least one of the momenta is far away from the Fermi surface.

\noindent {\em Pairing instability.--}
As $H_0$ can be written as non-interacting spinless fermions, it is not surprising that the pairing instability due to terms such as $f^\dagger_{\vec k_0/2+\vec k}f^\dagger_{\vec k_0/2-\vec k}$ is governed by the same type of logarithms which are responsible for $p$-wave superconductivity \cite{p-wave}. For small $U$, one can therefore expect that dimerization sets in below a transition temperature $T_c$ with
\begin{align}
T_c =E_{0,\alpha} \, e^{-c_\alpha {J }/{U}} \quad \text{for } U \ll J \label{Tc} \,,
\end{align}
where $E_{0,\alpha}$ is an energy scale of order $J$. The dimensionless constant $c_\alpha$ can be computed exactly from a one-loop renormalization group calculation or, alternatively, from a  BCS mean-field calculation. In the following we will use BCS theory to compute $T_c$ and the structure of the order parameter. Our approach allows to calculate $c_\alpha$ exactly, but unfortunately not the prefactor $E_{0,\alpha}$ \cite{FootnoteFluxGap}. 

To describe the dimerized phase, we consider the BCS-style Hamiltonian
\begin{align}
H_{\rm BCS}=\sum_k \epsilon_k f^\dagger_{\bf k} f^{\phantom\dagger}_{\bf k} +\sum_{k_x>0}\Delta_{\vec k}  f^\dagger_{\kk/2+\bf k} f^\dagger_{\kk/2-\bf k}+h.c. \,,
\end{align}
where the odd order parameter, $\Delta_{\vec k}=-\Delta_{-\vec k}$, is computed from the mean-field equation
\begin{align}
\Delta_{\vec{k}}=2 \sum_{\vec q}& V^1_{\kk/2+\vec k,\kk/2-\vec k,\kk/2+\vec q} \langle f_{\kk/2+\vec q}f_{\kk/2-\vec q} \rangle \label{BCS} \\
&+12\,  V^2_{\kk/2+\vec k,\kk/2-\vec k,\kk/2+\vec q} \langle f^\dagger_{\kk/2+\vec q}f^\dagger_{\kk/2-\vec q}\rangle \,,\nonumber
\end{align}
where we assumed that the interactions in Eq.~(\ref{Hlow}) have been completely antisymmetrized with respect to the fermionic operators. All expectation values are computed with $H_{\rm BCS}$. Note that in contrast to the standard BCS theory there is no $U(1)$ symmetry. It turns out that time-reversal symmetry (\ref{Tsym}) is {\em not} spontaneously broken which leads to a purely imaginary  $\Delta_{\vec k}$, $\Delta_{\vec k}^*=-\Delta_{\vec k}$. The latter, together with $\Delta_{\vec k}=-\Delta_{-\vec k}$,  already implies that pairing cannot gap the Fermi surface completely, but only reduce it to an odd number of lines. 

To simplify  Eq.~(\ref{BCS}) and to solve it with high numerical precision  to leading logarithmic order, we rewrite the $\vec q$ integration into an integral on the Fermi surface and an energy integration perpendicular to it, $\int d^3 \vec q = \int_{\rm FS} d^2 \vec k_F \int N_{\vec k_F}(\epsilon) d\epsilon$.  We furthermore approximate the directional dependent density of state by its value of the Fermi surface $N_{\vec k_F}(0)=\frac{1}{(2 \pi)^3 v_F(\vec k_F)}$ for $-2 J<\epsilon<2 J$. Similarly, we approximate both $\Delta_{\vec k}$ and the matrix elements of $V^1$ and $V^2$ by their values on the Fermi surface.  This allows for an accurate evaluation of the $\epsilon$ integration, $\int d\epsilon  \langle f_{\kk/2+\vec q}f_{\kk/2-\vec q} \rangle\approx  \Delta_{\vec q_F} \int_{0}^{2 J} d\epsilon \frac{\tanh \sqrt{\epsilon^2+|\Delta_{\vec q_F}|^2}/(2 T)}
{\sqrt{\epsilon^2+|\Delta_{\vec q_F}|^2}}$. The last term is evaluated numerically but it is also well described by $\Delta_{\vec q_F} \ln\!\left[4 J/\sqrt{|\Delta_{\vec q_F}|^2+(1.762\, T)^2}\right]$. 
For the plots of this paper, we discretize the Fermi surface using a total of 1344 patches.

\begin{figure}[t]
\begin{center}
\includegraphics[width=0.8 \linewidth]{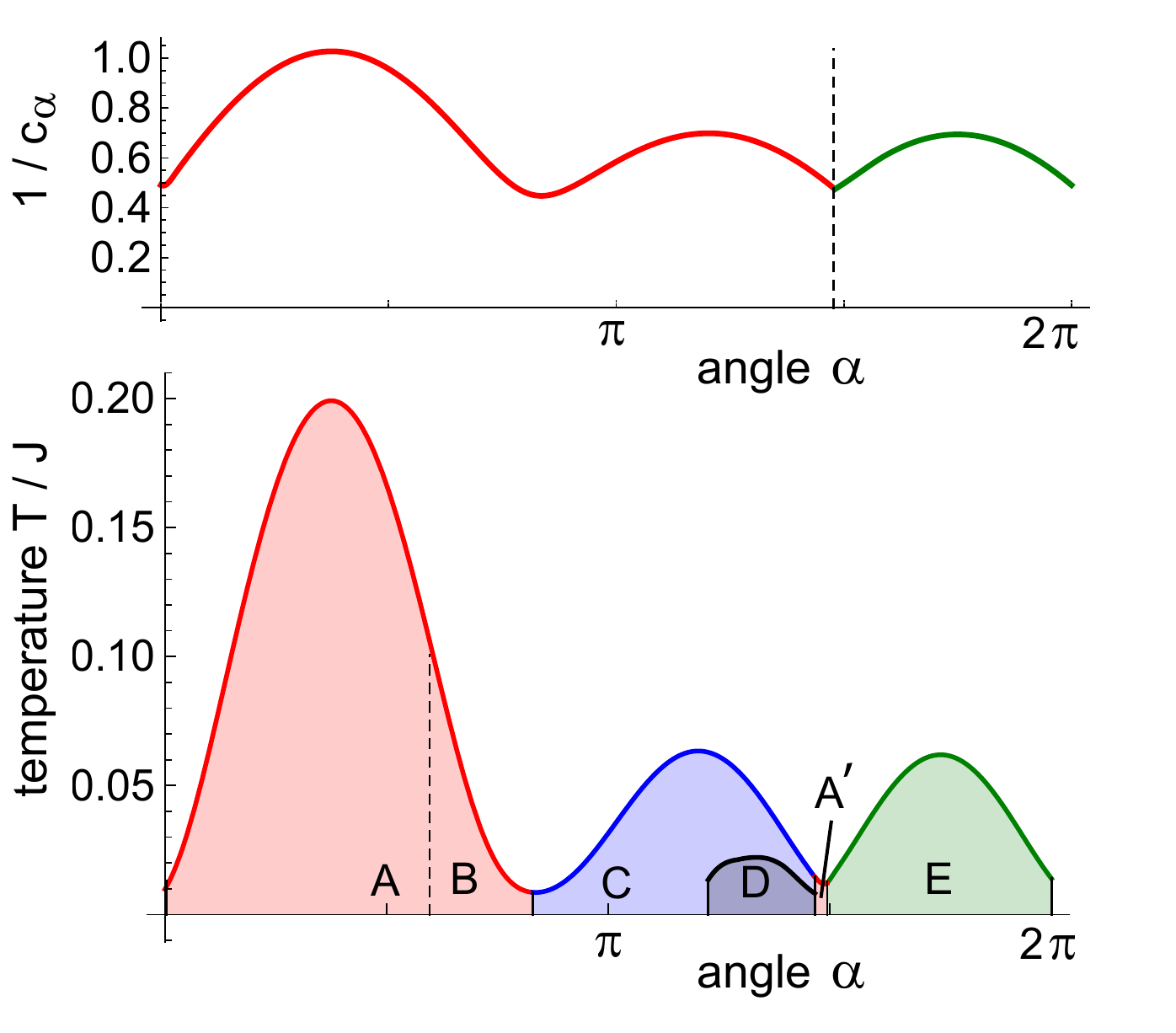}\\
\includegraphics[width=0.8 \linewidth]{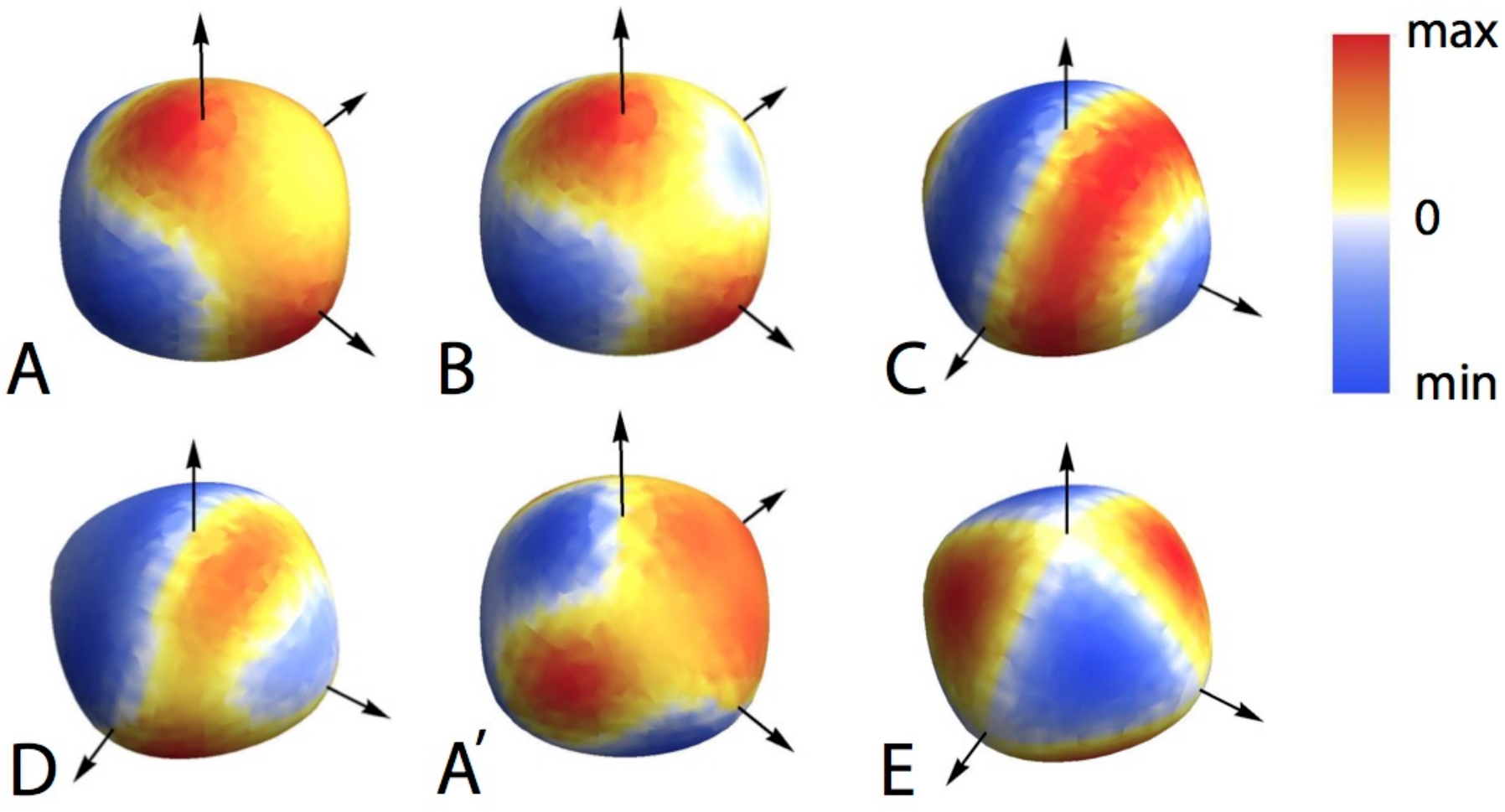}
\end{center}
\caption{(color online) Upper figure: The largest eigenvalue of the linearized BCS equation determines $c_\alpha$ in Eq.~(\ref{Tc}). Middle figure: Phase diagram for $U=0.4 J$ as function of the angle $\alpha$. There are 4 different phases distinguished by symmetry (see  table \ref{tab1}). Lower figure: For the five  phases A-E the angular distribution of the order parameter $\Delta_q$  on the Majorana surface is shown ($\alpha=0.5 \pi, 0.7 \pi, \pi, 1.35 \pi, 1.48 \pi, 1.75 \pi$). While there is a single line node in phase A, there are three non-crossing line nodes for B, C, D, and three line nodes with 6 crossing points for E.}
\label{fig:PD}
\end{figure}

\noindent {\em Phase diagram.--}
Using the largest eigenvalue of the linearized BCS equation (\ref{BCS}), one can directly determine $c_\alpha$ in Eq.~(\ref{Tc})  with $E_{0,\alpha}\approx 2.27 J$ within our cutoff scheme. $1/c_\alpha$ is plotted in the upper panel of Fig.~\ref{fig:PD}. For $0<\alpha<4.65$ (red) the eigenvalue is three-fold degenerate. The three eigenvectors form an irreducible representation of the point group $O$ and transform like $k_x,k_y$ and $k_z$ ($T_1$ representation, see supplemental material). For $4.65<\alpha<2 \pi$ (green), in contrast, the eigenvalue is unique and the eigenvector transforms like $k_x k_y k_z$ ($A_2$ representation). Analyzing the BCS equation beyond the linear approximation, one obtains the phase diagram shown for $U=0.4 J$ in the middle panel of Fig.~\ref{fig:PD} as function of temperature $T$ and $\alpha$. For other values of $U$ one obtains very similar phase diagrams, but all transition temperatures and the ratio of the largest and smallest $T_c$ change exponentially according to Eq.~(\ref{Tc}).

\begin{table}[t]
\vspace*{0.5cm}
\begin{tabular}{c|c|c|c|c}
 & order parameter &  point group& domains & line nodes\\  \hline 
  \begin{tabular}{cc}  A, A'\\ B \end{tabular} & $k_x+k_y+k_y$ & D$_3$ (32) & 8 & \begin{tabular}{cc} 1 \\ 3 \end{tabular}\\ \hline
C&$k_x$ & D$_4$ (422) & 6 & 3   \\ \hline
D&$c_1 k_x + c_2 k_y$ \ & C$_2$ (2) & 24 & 3   \\ \hline
E& $k_x  k_y k_z$ & O (432) & 2 & 3 (crossing)  \end{tabular}
\caption{Overview  of the symmetry broken phases, see Fig.~\ref{fig:PD}. In all phases the unit cell is doubled. The first column shows schematically the symmetry of the order parameter (momenta measured from $k_0/2$). Furthermore,  the point group of the symmetry broken phase, the number of domains and the number of line nodes  are tabulated.\label{tab1}}
\end{table}

 We find five different phases, denoted $A-E$ in Fig.~\ref{fig:PD}, which are distinguished by symmetry and/or the number of line nodes; the corresponding distribution of order parameters is shown in the lower panel of Fig.~\ref{fig:PD}. 
In table~\ref{tab1} an overview of the properties of the symmetry-broken phases is given.
In all phases the unit-cell has doubled. In phase $E$ this is the only broken symmetry.
In the phases $A-D$ besides the translational symmetry also other lattice symmetries are broken. As at $T_c$ the eigenvalue $c_\alpha$ is  three-fold degenerate, the phase diagram can be understood best in terms of a three-component order parameter, $\boldsymbol \Phi=(\Phi_1,\Phi_2,\Phi_3)$ where $\Phi_i$ corresponds to an eigenvector transforming like $k_i$. Close to $T_c$, one can use a Landau description of the free energy of the form 
\begin{equation} 
F\approx (T-T_c) \boldsymbol \Phi^2+g (\boldsymbol \Phi^2)^2+g'  (\Phi_1^4+\Phi_2^4+\Phi_3^4)+O(\Phi^6) \label{GL} \,.
\end{equation}
It predicts that just below $T_c$ the order parameter points for $g'>0$ in one of the $(111)$ directions (phases A, A', and B colored in red in Fig.~\ref{fig:PD}). The order parameter therefore has a 120\degree-rotation symmetry around this axis, see also the lower panel in Fig.~\ref{fig:PD}. In the crystal structure this $(111)$ direction will be shortened or elongated (depending on the sign of the coupling to the lattice).
In total the number of domains is 8, corresponding to four possible structural transformations  times two different translational domains.

For $g'<0$, in contrast, the order parameter points along a $(100)$ direction (phase C with 6 domains) and the lattice will shrink or expand along that direction. Interestingly, it turns out that all phases besides the phase $A$ are characterized by three line nodes instead of a single one, see Fig.~\ref{fig:PD}. In the language of superconductivity, one would call this an ``extended $p$-wave order parameter'' \cite{extended-p-wave}.
The simple Landau theory (\ref{GL}) can fail at low $T$ and indeed phase $D$ cannot be described by terms up to order $\Phi^4$. Here further symmetries are broken (see table \ref{tab1}), $\boldsymbol \Phi$ points in a low-symmetry direction, e.g., $(c_1, c_2, 0)$, and the ordered phase is characterized by 24 different domains. In this case the only remaining symmetry of the order parameter is that it changes sign by 180\degree-rotation around the (001) axis.

\noindent{\em Experimental signatures.--}
The spin-Peierls instability of 3D Kitaev spin liquids with a Majorana Fermi surface  and its resulting nodal-line QSL provide a distinct experimental fingerprint, which could facilitate the ongoing search for Kitaev spin liquids in spin-orbit entangled $j=1/2$ Mott insulators.
In thermodynamics, one expects to observe 
(i) the absence of magnetic order breaking time-reversal symmetry,
(ii) an approximately constant specific heat coefficient $c_v/T$ over a wide intermediate temperature range (indicative of the Majorana Fermi surface),  
(iii) a structural phase transition with unit-cell doubling at small temperatures $T_c$, 
and (iv) a low-temperature specific heat with $c_v/T \approx T$  (indicative of the nodal lines) with possible logarithmic corrections (arising from the crossing points of the nodal lines for phase $E$). Besides the structural phase transition discussed in this publication, three dimensional $\mathbb Z_2$ gauge theories (without monopoles \cite{hermanns14}) also show a finite-$T$ phase transition  \cite{finiteTGauge1,finiteTGauge2} of (inverted) Ising universality class. Note that we have implicitly assumed in our calculation that the $T_c$ of the structural transition is well below this transition temperature where flux lines proliferate.

\noindent {\em Summary.--}
We have shown that a time-reversal symmetric spin liquid with Majorana surface is always unstable and  spontaneously undergoes a spin-Peierls transition to a nodal QSL at low temperatures. We expect that this is a generic property of Majorana surfaces. To show this, it is useful to classify time-reversal invariant Majorana systems by the value of the vector $\vec{k}_0$
characterizing the time-reversal operation in Eq.~\eqref{eq:T} or, equivalently, Eq.~\eqref{Tsym}.
If $\vec{k_0}$ vanishes, terms of the form $f^\dagger_{\vec k_0/2+\vec k} f^\dagger_{\vec k_0/2-\vec k}$ occur even in the absence of symmetry breaking and -- instead of a Fermi surface -- only a state with nodal lines forms \cite{Mandal09,finiteTGauge2}. Fermi surfaces therefore exist only for finite $\vec k_0$. In this case, however, time-reversal invariance guarantees the existence of a BCS-type instability. As any interaction of Majorana excitations always involves four different sites, the momentum-dependent interaction will always have
an attractive channel leading to a transition where Majorana pairs condensate at finite momentum $\vec k_0$.
We therefore expect that Majorana surfaces can only survive for $T \to 0$ in cases where  time-reversal symmetry is broken either spontaneously or explicitly, e.g. by an external magnetic field -- a situation which we plan to investigate in the future.

\begin{acknowledgments}
\noindent {\em Acknowledgments.--}
M.H. acknowledges partial support through the Emmy-Noether program of the DFG.
\end{acknowledgments}

\onecolumngrid

\begin{appendix}

\section{Symmetry transformations}

Following Kitaev's original approach \cite{kitaev2006} we can solve the Hamiltonian \eqref{eq:Kitaev} by representing the spins in terms of four Majorana fermions 
\begin{align}
\sigma_j^\alpha=i a_j^\alpha c_j
\end{align}
and writing \eqref{eq:Kitaev} 
\begin{align}
  H_{\rm Kitaev} = J \sum_{\gamma \rm-links} \hat{u}_{j,j'}^\gamma i c_j c_j' \,,
\end{align}
where we defined bond operators $\hat u_{j,j'}^\gamma=i a_j^\gamma a_{j'}^\gamma$. By mapping the spin degree of freedom to four Majorana fermions, we introduced a $\mathbb Z_2$ gauge degree of freedom on each bond and the physical space is defined by $a_j^x a_j^y a_j ^z c_j |\mbox{phys}\rangle=|\mbox{phys}\rangle$. Following Kitaev, we fix a gauge to do our computations, see Ref. \cite{hermanns14} for details on the gauge choice. 
As a result, lattice symmetries as well as time-reversal symmetry may have to be supplemented with a gauge transformation in order to be a symmetry of the gauge-fixed Hamiltonian. 
Table \ref{tab:symmetries} lists the appropriate gauge transformation for each of the 48 symmetry transformations of the $I4_132$ space group, implemented by
\begin{align}
c_\alpha (\vec R)  \rightarrow c_\alpha (\vec R) \sigma_\alpha \delta^{\vec k_0 R}.
\end{align}
Here $\sigma_j$, $j=1,\ldots,4$ denote the sign factor for the Majorana operator at position $j$ -- listed in the second column of Table \ref{tab:symmetries}.  The values of $\delta=\pm 1$ is denoted in the third column. 

\begin{table*}[h!]
\begin{tabular}{ccc|rrrr|r||ccc|rrrr|r}
\toprule
\multicolumn{3}{c}{coordinate}& \multicolumn{4}{c}{gauge transformations}&\multicolumn{1}{c}{staggered}&
\multicolumn{3}{c}{coordinate}& \multicolumn{4}{c}{gauge transformations}&\multicolumn{1}{c}{staggered}\\
\bottomrule
\hline \hline
 x& y& z& 1& 1& 1& 1 & 1 &  1/2 + x & 1/2 + y & 1/2 + z & 1& 1& 1& 1 & 1\\
 1/2 - x& -y& 1/2 + z & 1& 1& -1& -1 & 1&  -x& 1/2 - y& z & 1& 1& -1& -1 & 1\\
 -x& 1/2 + y& 1/2 - z & 1& -1& 1& -1 & 1&  1/2 - x& y& -z & 1& -1& 1& -1 & 1\\
 1/2 + x& 1/2 - y& -z & 1& -1& -1& 1 & 1&  x& -y& 1/2 - z & 1& -1& -1& 1 & 1\\
 z& x& y & 1& 1& 1& 1 & 1&  1/2 + z& 1/2 + x& 1/2 + y & 1& 1& 1& 1 & 1\\
 1/2 + z& 1/2 - x& -y & 1& -1& -1& 1 & 1&  z& -x& 1/2 - y & 1& -1& -1& 1 & 1\\
 1/2 - z& -x& 1/2 + y & 1& 1& -1& -1 & 1&  -z& 1/2 - x& y & 1& 1& -1& -1 & 1\\
 -z& 1/2 + x& 1/2 - y & 1& -1& 1& -1 & 1&  1/2 - z& x& -y & 1& -1& 1& -1 & 1\\
 y& z& x & 1& 1& 1& 1 & 1&  1/2 + y& 1/2 + z& 1/2 + x & 1& 1& 1& 1 & 1\\
 -y& 1/2 + z& 1/2 - x & 1& -1& 1& -1 & 1& 1/2 - y& z& -x & 1& -1& 1& -1 & 1\\
 1/2 + y& 1/2 - z& -x & 1& -1& -1& 1 & 1&  y& -z& 1/2 - x & 1& -1& -1& 1 & 1\\
 1/2 - y& -z& 1/2 + x & 1& 1& -1& -1 & 1&  -y& 1/2 - z& x & 1& 1& -1& -1 & 1\\
 \hline
 3/4 + y& 1/4 + x& 1/4 - z & 1& -1& -1& -1 & -1&   1/4 + y& 3/4 + x& 3/4 - z & 1& -1& -1& -1 & -1\\
 3/4 - y& 3/4 - x& 3/4 - z & 1& 1& -1& 1 & -1& 1/4 - y& 1/4 - x& 1/4 - z & 1& 1& -1& 1 & -1\\
 1/4 + y& 1/4 - x& 3/4 + z & 1& -1& 1& 1 & -1& 3/4 + y& 3/4 - x& 1/4 + z & 1& -1& 1& 1 & -1\\
 1/4 - y& 3/4 + x& 1/4 + z & 1& 1& 1& -1 & -1& 3/4 - y& 1/4 + x& 3/4 + z & 1& 1& 1& -1 & -1\\
 3/4 + x& 1/4 + z& 1/4 - y & 1& -1& -1& -1 & -1& 1/4 + x& 3/4 + z& 3/4 - y & 1& -1& -1& -1 & -1\\
 1/4 - x& 3/4 + z& 1/4 + y & 1& 1& 1& -1 & -1& 3/4 - x& 1/4 + z& 3/4 + y & 1& 1& 1& -1 & -1\\
 3/4 - x& 3/4 - z& 3/4 - y & 1& 1& -1& 1 & -1& 1/4 - x& 1/4 - z& 1/4 - y & 1& 1& -1& 1 & -1\\
 1/4 + x& 1/4 - z& 3/4 + y & 1& -1& 1& 1 & -1& 3/4 + x& 3/4 - z& 1/4 + y & 1& -1& 1& 1 & -1\\
 3/4 + z& 1/4 + y& 1/4 - x & 1& -1& -1& -1 & -1& 1/4 + z& 3/4 + y& 3/4 - x & 1& -1& -1& -1 & -1\\
 1/4 + z& 1/4 - y& 3/4 + x & 1& -1& 1& 1 & -1& 3/4 + z& 3/4 - y& 1/4 + x & 1& -1& 1& 1 & -1\\
 1/4 - z& 3/4 + y& 1/4 + x & 1& 1& 1& -1 & -1& 3/4 - z& 1/4 + y& 3/4 + x & 1& 1& 1& -1 & -1\\
 3/4 - z& 3/4 - y& 3/4 - x & 1& 1& -1& 1 & -1&  1/4 - z& 1/4 - y& 1/4 - x & 1& 1& -1& 1 & -1\\
 \hline\hline
\end{tabular}
\caption{Overview of the 48 symmetry transformations of the I4$_1$32 symmetry group of the hyperoctagon lattice and the associated gauge transformation in the  Majorana model. The left column displays the symmetry transformation in real space. The second column displays which sites within the unit cell acquire a sign. A `-1' in the third column indicates, that the gauge transformation is `staggered' for neighboring unit cells, i.e. it contains an additional factor  $(-1)^{\vec k_0 \vec R}$. }
\label{tab:symmetries}
\end{table*}

\newpage
\section{Character table for the point group $O$}

\begin{table*}[h!]
\begin{tabular}{c|c|c|c|c|c|c|c|c}
O&E&8C$_3$&6C$_2$'&6C$_4$&3C$_2$&linear functions&quadratic functions &cubic functions\\
\hline
A$_1$ & 1&1&1&1&1&-&$x^2+y^2+z^2$&-\\
A$_2$&1&1&-1&-1&1&-&-&xyz\\
E&2&-1&0&0&2&-&$(x^2-y^2,2z^2-x^2-y^2)$&-\\
T$_1$&3&0&-1&1&-1&$(x,y,z)(R_x,R_y,R_z)$&-&$(x^3,y^3,z^3)[x(z^2+y^2),y(z^2+x^2),z(x^2+y^2)]$\\
T$_2$&3&0&1&-1&-1&-&(xy,xz,yz)&$[x(z^2-y^2),y(z^2-x^2),z(x^2-y^2)]$
\end{tabular}
\caption{Character table for the point group $O$. }
\label{tab:symmetries}
\end{table*}

\end{appendix} 


\begin{thebibliography}{99}
\bibitem{FootnotePairing}
Note that for other three-dimensional, tricoordinated lattices (such as the hyperhoneycomb lattice) $\kk$ actually vanishes. For these lattices, these pairing terms are expected to occur even in the absence of spontaneous symmetry breaking. As a consequence, none of the QSL ground states of these lattices exhibit a Majorana Fermi surface but a nodal line structure instead \cite{Mandal09}.
\bibitem{FootnoteThirdTerm}
A third type of interaction term involving a Majorana and its three neighboring sites (e.g., $c_1(\mathbf R)c_2 (\mathbf R)c_3(\mathbf R)c_4(\mathbf R-\mathbf a_3)$) turns out to be forbidden by time reversal symmetry.
\bibitem{FootnoteFluxGap}
For our calculation we assume that $T_c$ remains well below the gap for flux-loop excitations, which we do not include in our treatment. The numerical value of the flux gap is $\approx 0.18 J$ for the hyperoctagon model at hand. Note that some of the $T_c$ values in the middle panel of Fig.~\ref{fig:PD} overshoot this value due to the relatively large value of $U$ chosen for this illustration.
\end{thebibliography}
\end{document}